\documentclass[a4paper,11pt]{article}
\usepackage{pos}
\usepackage{subcaption} 
\usepackage{hyperref}
\usepackage{setspace}

\title{Design and performance of the prototype Schwarzschild-Couder Telescope camera}
 \ShortTitle{Design and performance of the pSCT Camera}

\author*[a]{L.~P.~Taylor}
\affiliation[a]{Department of Physics and Wisconsin IceCube Particle Astrophysics Center,\\ University of Wisconsin, Madison, WI 53706, USA}
\forColl{CTA SCT} 
\emailAdd{ltaylor23@wisc.edu}

\abstract{The Cherenkov Telescope Array (CTA) is the next-generation ground-based observatory for very-high-energy gamma-ray astronomy. An innovative 9.7 m aperture, dual-mirror Schwarzschild-Couder Telescope (SCT) design is a candidate design for CTA Medium-Sized Telescopes. A prototype SCT (pSCT) has been constructed at the Fred Lawrence Whipple Observatory in Arizona, USA. Its camera is currently partially instrumented with 1600 pixels covering a field of view of 2.7 degrees square. The small plate scale of the optical system allows densely packed silicon photomultipliers to be used, which combined with high-density trigger and waveform readout electronics enable the high-resolution camera. The camera's electronics are capable of imaging air shower development at a rate of one billion samples per second. We describe the commissioning and performance of the pSCT camera, including trigger and waveform readout performance, calibration, and absolute GPS time stamping. We also present the upgrade to the camera, which is currently underway. The upgrade will fully populate the focal plane, increasing the field of view to 8 degree diameter, and lower the front-end electronics noise, enabling a lower trigger threshold and improved reconstruction and background rejection.}

\FullConference{37$^{\rm{th}}$ International Cosmic Ray Conference (ICRC 2021)\\July 12th -- 23rd, 2021\\Online -- Berlin, Germany}

\begin{document}
\maketitle

\section{Introduction}

\begin{figure}
    \centering
    \includegraphics[width=0.95\textwidth]{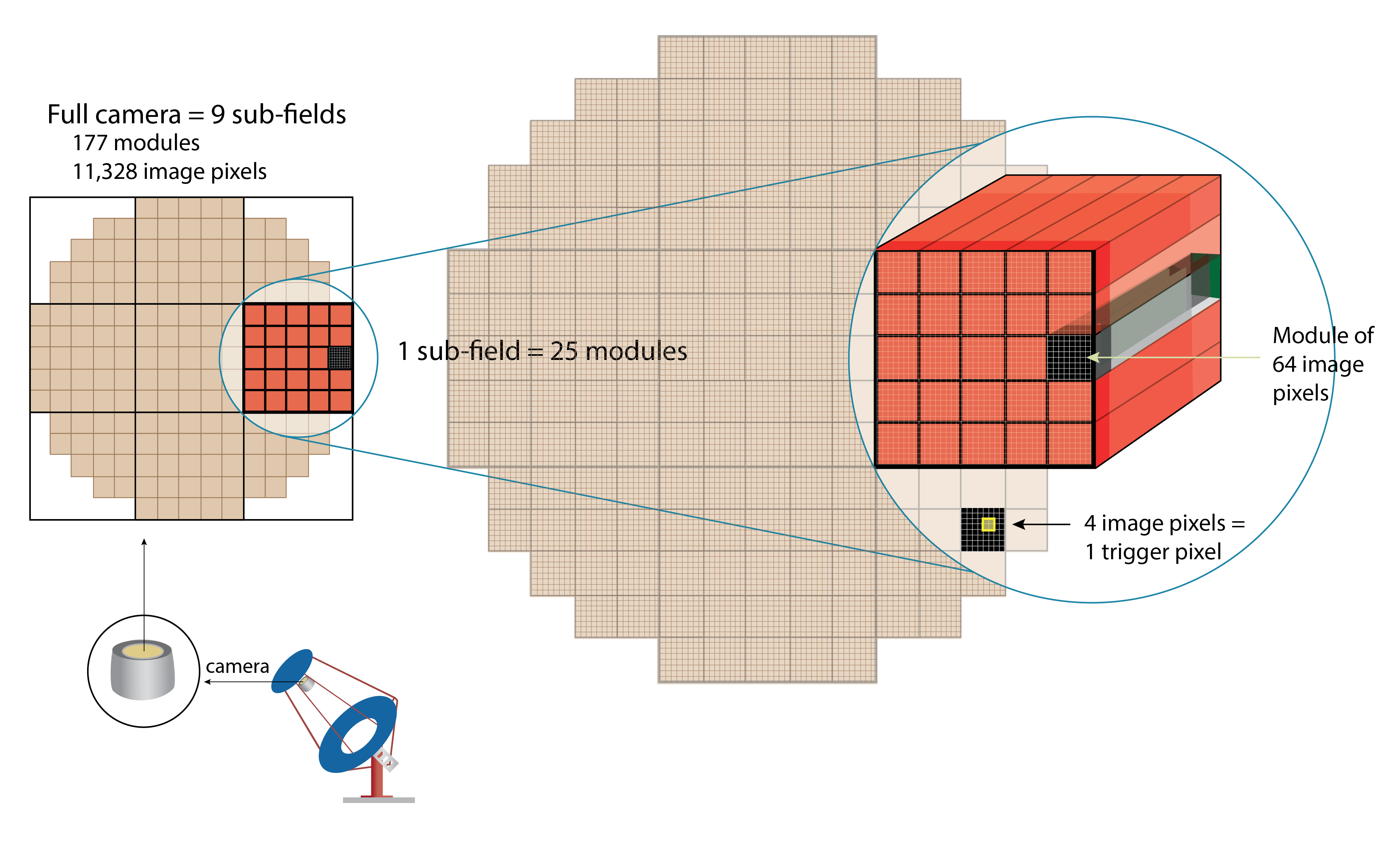}
    \caption{The full pSCT camera is comprised of 177 modules divided into 9 sectors. Each sector can hold up to 25 modules with the sectors in each corner containing fewer modules such that the final focal plane will be roughly circular. Each module contains 64 image pixels with the full camera containing 11,328 pixels. Currently, only the center sector is instrumented. 24 modules are installed, with the central module not yet installed to accommodate optical alignment hardware.}
    \label{fig:CameraDesign}
\end{figure}

Earth's atmosphere is opaque to very-high-energy (VHE) photons, however these photons initiate air showers upon hitting the atmosphere. The shower constituents move faster than the speed of light in the atmosphere, resulting in Cherenkov radiation which can be detected at the ground by Imaging atmospheric Cherenkov Telescopes (IACTs). The Cherenkov Telescope Array (CTA) is an IACT array with one site in each hemisphere. CTA will feature three telescope sizes and cover energies between 20 GeV and 300 TeV \cite{acharyya2019monte}. Most IACTs use Davies-Cotton (single-mirror) optics. However, the prototype Schwarzschild-Couder Telescope (pSCT), a candidate for the CTA Medium Sized Telescope, will use dual-mirror optics. The Schwarzschild-Couder optics produce a wide field of view, an excellent optical point spread function, and a smaller plate scale when compared to single-mirror designs \cite{vassiliev2008schwarzschild, deivid}.

This small plate scale allows the use of silicon photomultipliers (SiPMs) in place of traditional photomultiplier tubes. SiPMs are smaller than traditional PMTs enabling improved image resolution. The current pSCT is instrumented with 25 modules (1600 pixels) and has a 2.7$^{\circ}$ field of view. An upgrade to the pSCT is underway which is expected to be complete by the end of 2022. The upgrade will include improvements to the camera consisting of an update of the camera module front-end electronics (FEE), the camera backend electronics, and the module SiPMs as well as fully populating the focal plane (increasing the number of pixels to 11,328 and the field of view to 8$^{\circ}$ diameter.

\section{Commissioning and Performance}

Commissioning of the pSCT camera is an ongoing process and in the last two years, significant improvements have been made in the quality of pSCT data taking. Among these are validation of the trigger path mapping, increasing the trigger path dynamic range, and the inclusion of all modules in the trigger. Concurrent with commissioning work, initial observations of the Crab Nebula were made. The observation and detection of the Crab Nebula by the pSCT are described in detail in \cite{brent, adams2021detection}.

\subsection{Validation of the Trigger Path}
\label{Trigger Mapping}

\begin{figure}
    \centering
    \includegraphics*[clip, trim={0.1cm 0.1cm 0.1cm 0.1cm}, width=0.8\textwidth]{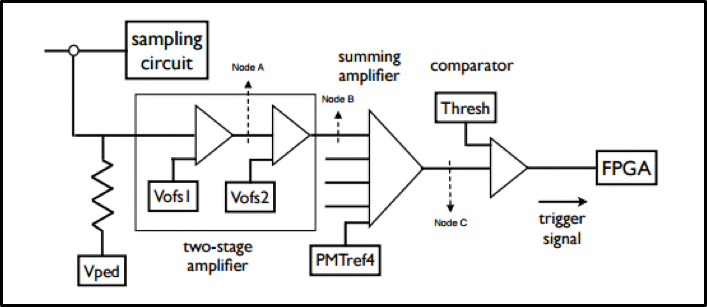}
    \caption{Circuit diagram for one trigger pixel. \cite{Wachtendonk2018} In TARGET7, four (2x2) image pixels are amplified individually via two inverting amplifiers (with Vofs1 and Vofs2 as reference) and then summed with an inverting summing amplifier (with PMTref4 as reference) to produce the signal for one trigger pixel. The output of the summing inverting amplifier (Node C) originally had a desired set-point of 1.25~V at baseline. This desired set-point is achieved by tuning the reference values Vofs1, Vofs2, and PMTref4 such that Node C reaches this value for a baseline signal. This desired set-point was changed to 2.1~V resulting in a greater dynamic range for our expected signal (downgoing at Node C). Note that due to inversion of the signal along the amplifier chain, a higher Thresh value corresponds to a lower physical threshold.}
    \label{fig:CircuitDiagram}
\end{figure}
    
\begin{figure}
    \centering
    \begin{subfigure}{.49\textwidth}
      \centering
      \includegraphics[clip, trim={0cm 19.55cm 0cm 2cm}, width=0.8571\textwidth]{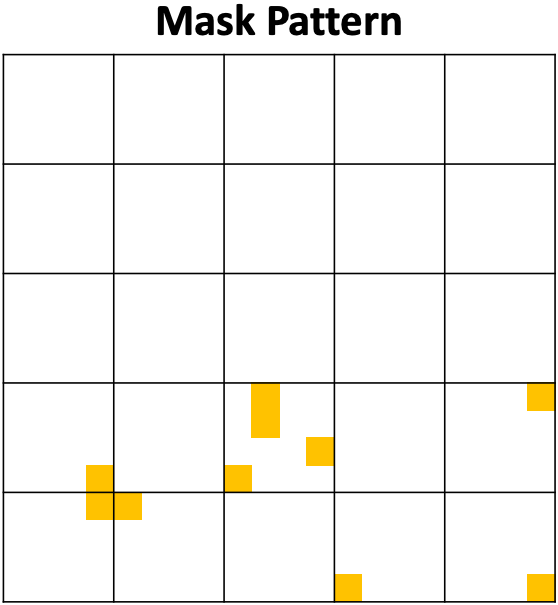}
      \includegraphics[width=0.844\textwidth]{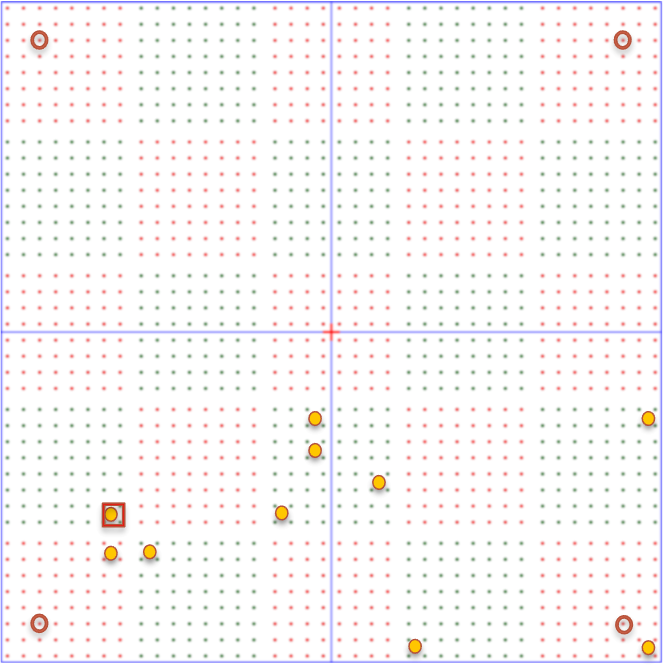}
      \caption{Design of the physical mask. Each module is represented by a group of 64 (8x8) image pixels (dots). Red empty circles highlight the image pixels used to align the mask using the data path image.  Yellow filled circles are the trigger pixels used for checking the trigger pattern. Only one group of trigger pixels is capable of satisfying the three-fold coincidence required to produce a backplane trigger.}
      \label{fig:MaskPattern}
    \end{subfigure}
    \hfill
    \begin{subfigure}{.49\textwidth}
      \centering
      \includegraphics[clip, trim={0cm 0cm 0cm 2cm}, width=0.9143\textwidth]{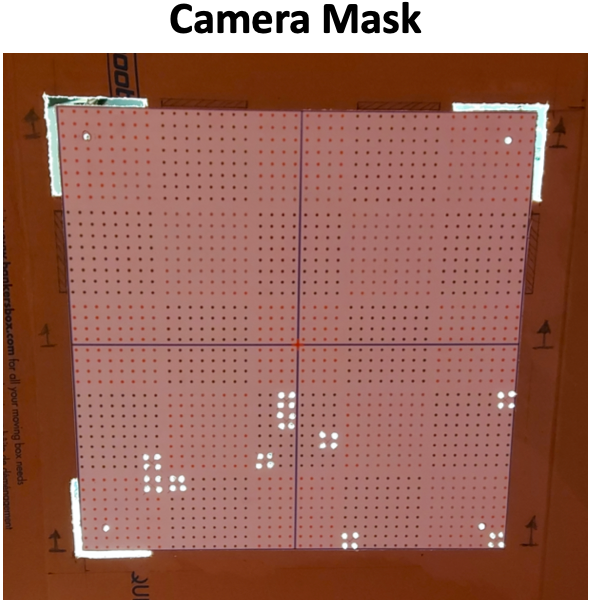}
      \caption{Realization of the physical mask used to cover the focal plane. Constructed from Cardboard and illuminated from behind. Each dot is the center of a 6x6 $\mathrm{mm}^2$ image pixel; 4 image pixels make up a trigger pixel. The corner slits are used for visual alignment, while the single-pixel openings in each corner aid with alignment using the data path. Ten trigger pixels were chosen to remain uncovered.}
      \label{fig:MaskPicture}
    \end{subfigure}
    \par\bigskip
    \begin{subfigure}{.49\textwidth}
      \centering
      \includegraphics[clip, trim={0cm 0cm 0cm 2cm}, width=\textwidth]{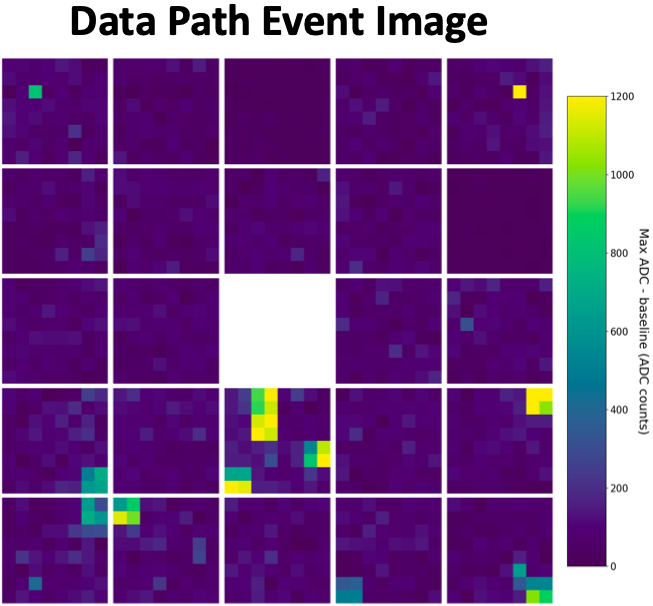}
      \caption{Data-path image taken with camera mask installed on the camera. The color scale is proportional to the signal amplitude in each pixel. Unmasked pixels are clearly visible in yellow/green and the reference pixels in each corner are well aligned. All pixels visible in the data path are also visible in the trigger hit map in Figure \ref{fig:Eventhit map}.}
      \label{fig:EventImage}
    \end{subfigure}
    \hfill
    \begin{subfigure}{.49\textwidth}
      \centering
      \includegraphics[clip, trim={0cm 0cm 0cm 2cm}, width=0.8571\textwidth]{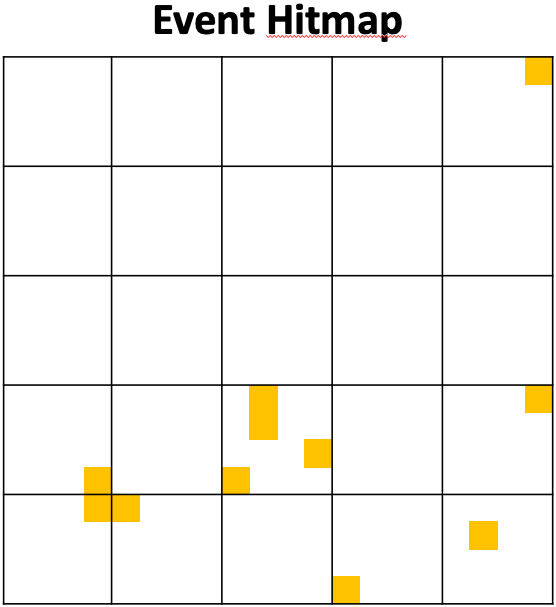}
      \caption{A binary image of module triggers (hit map). A yellow square indicates the pixel has triggered while a white square indicates that it has not. The hit map was read out concurrently with the event image in Figure \ref{fig:EventImage}. Only the cluster of three trigger pixels on the left side passes the 3-fold coincidence logic and produces a backplane trigger.}
      \label{fig:Eventhit map}
    \end{subfigure}
    \caption{An asymmetric cardboard mask was constructed and attached to the camera focal plane. The mask was designed such that there was only one grouping of pixels capable of producing a camera trigger (which uses three-fold-coincidence). A moderate intensity flasher was used to illuminate the camera through the mask. Event images from the data path and hit maps from the trigger path both match the mask pattern.}
    \label{fig:MaskTrigger}
\end{figure}

The pSCT trigger system contains two trigger levels: \textit{module triggers} and \textit{backplane triggers} (when integrated into an array a third \textit{telescope trigger} will also be included). The module level trigger system (Figure \ref{fig:CircuitDiagram}) monitors groups of four (2x2) image pixels (called a trigger pixel) for a threshold (Thresh) crossing of their analog sum. If this value is crossed for any trigger pixel a module trigger is sent to the backplane. The backplane then collects these module trigger signals from each of the camera modules. The FEE programmable pulse width (currently set to $\sim$10~ns) determines the coincidence resolving time of the pattern trigger. A module trigger overlap of approximately 1.5~ns is sufficient to generate a coincidence.

The backplane then forms backplane level triggers via a single trigger FPGA to form coincidences from the 400 module trigger inputs (16 module trigger inputs for each of the 25 modules). When three adjacent trigger pixels produce a module trigger at the same time, a backplane trigger is produced, prompting the camera to read out the event. A pixel is considered adjacent if it is orthogonal-adjacent or diagonal-adjacent (i.e. a pixel can have up to eight adjacent pixels). A timestamp along with a hit map (a binary image of module triggers) is attached to the backplane trigger.

In order to validate this trigger system, a physical mask was constructed out of cardboard and affixed to the camera focal plane. The mask was aligned visually with the focal plane using three slits at the corners of the focal plane. Four small holes were placed at each corner for data-path alignment. The mask was designed such that ten trigger pixels would be illuminated in an asymmetric pattern so that any mapping irregularities would be easy to spot. It was also designed such that only one grouping of trigger pixels would produce a backplane trigger. This grouping contained trigger pixels from three different modules in order to ensure that cross-module triggering was effective.

The camera was then illuminated by an LED flasher mounted on the secondary mirror (running at 10 Hz) and allowed to trigger using the usual three-fold coincidence. hit maps were read out concurrently. Data path images and hit maps were then compared to the mask pattern. Figure \ref{fig:MaskTrigger} shows a single example. The comparison showes that the data path mapping and trigger path mapping are consistent both with each other and the physical layout of the camera.

\subsection{Improving Dynamic Range}
\label{Data Path Dynamic Range}

\begin{figure}
    \centering
    \includegraphics[width=0.7\textwidth]{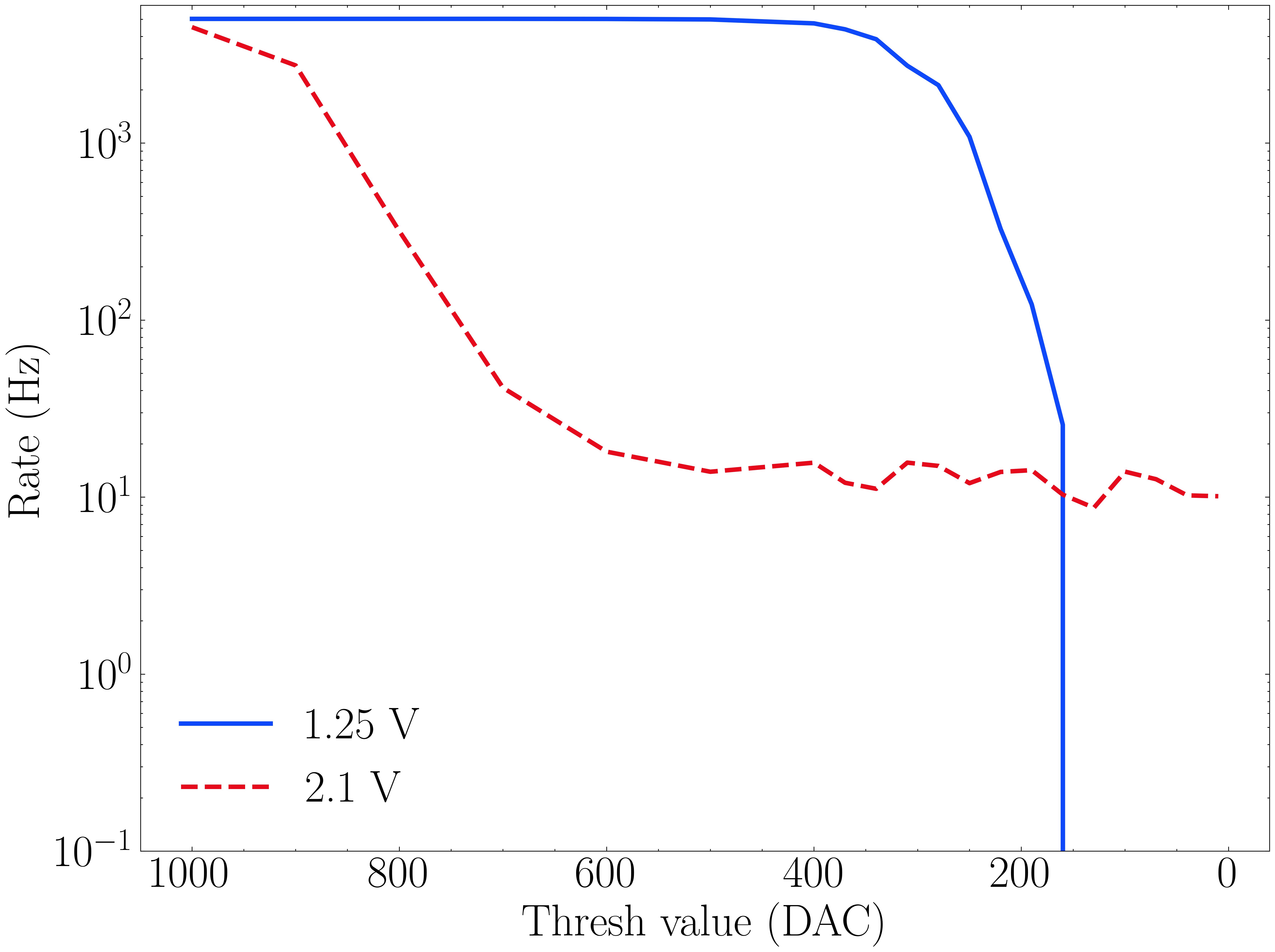}
    \caption{Two rate scans, comparing Thresh value to the trigger rate, each with all modules included in the trigger and a flasher running at 10 Hz at constant intensity. The baseline value of the output of the summing inverting amplifier (Node C) shown in Figure \ref{fig:CircuitDiagram} has been changed from 1.25~V to 2.1~V in order to improve the dynamic range of our expected signal. The upper (blue) rate scan was taken with the previous value of 1.25~V. The lower (red) rate scan was taken with a new value of 2.1~V. The new configuration shows a significant improvement in the dynamic range.}
    \label{fig:RateScanPMTref4}
\end{figure}
    
Modules contain 64 image pixels. In the trigger path these image pixels are grouped into trigger pixels before trigger logic is applied. A trigger pixel is comprised of the analog sum of four image pixels. The individual signals are combined using a summing amplifier with the value PMTref4 used as reference (see Figure \ref{fig:CircuitDiagram}). Nominally, the output of the summing amplifier (Node C) can vary between 0--2.5~V. However, the range of linearity is 0.4--2.1~V. The output of this amplifier is fed into a comparator where the signal is compared with the Thresh value to determine if there is a trigger. Due to inversions along the amplifier chain a high Thresh value corresponds to a low physical threshold.

Originally, the Node C was selected to be 1.25~V at baseline, the midpoint of the possible range. This desired output was achieved by tuning the reference values Vofs1, Vofs2, and PMTref4 for a baseline input. Using the midpoint value meant that both positive and negative signals had the same dynamic range. However, pulses coming into the inverting summing amplifier are unidirectional, meaning half of this dynamic range was not being used. For this reason we decided to raise the baseline value of Node C to 2.1~V as that was the highest value at which the summing amplifier maintained linearity. This change effectively extended the trigger dynamic range for our expected signal. Two rate scans (comparing trigger rate vs Thresh value) were taken in order to illustrate this effect. Both rate scans were taken with the flasher running at 100 Hz. One was taken with the old settings but with all modules in the trigger (when typically only US modules would be included - See section \ref{Including all modules in the trigger}) and one with the new settings. Figure \ref{fig:RateScanPMTref4} shows these rate scans. Prior to the adjustment the camera exhibited a very narrow trigger dynamic range, quickly entering the baseline noise. The new setting improved the dynamic range significantly, clearly showing the flasher plateau.

\subsection{Including All Modules in the Trigger}
\label{Including all modules in the trigger}

Currently the camera includes two types of SiPMs. Fifteen of the modules (US) use Hamamatsu SiPMs (model S12642-0404PA-50(X)) while the remaining nine (INFN) use FBK NUV-HD SiPMs. \cite{gola2019nuv} INFN modules have approximately twice the gain of US modules, due to a two-times larger FEE electronics gain combined with different SiPM gains. Because of this the conversion between Thresh value and photoelectrons for both modules is different. This difference in threshold meant that for the same Thresh value INFN modules would dominate the trigger. This fact combined with the geometry of the camera, which houses INFN modules on the edge of the focal plane, meant that it was beneficial to exclude INFN modules from the trigger during Crab observations. In an effort to update the trigger the Thresh to photoelectron conversion was found for both US and INFN modules with the intention of selecting two Thresh values (one for US and one for INFN modules) which corresponded to the same photoelectron threshold. However, there is no threshold at which both US and INFN modules can be included in the trigger. Rather than continue to leave the INFN modules out of the trigger completely, it was decided that they would be included at a thresh value which keeps them subdominant to the US modules (approximately one tenth the trigger rate). The INFN modules were included in the trigger starting March 17, 2021.

\subsection{Temperature Dependent Trigger Tuning}
\label{Temperature Dependent Trigger Tuning}

At the module level, the trigger path consists of two inverting amplifiers and an inverting summing amplifier (Figure \ref{fig:CircuitDiagram}). Node C is desired to be 2.1~V when the baseline of the signal is at its nominal value (see Section \ref{Data Path Dynamic Range}). This output depends on the values Vofs1, Vofs2, and PMTref4, which are the reference values for the two inverting amplifiers and the inverting summing amplifier, respectively. Due to slight differences in component values between circuits, these values must be tuned to reach the desired output. Originally, this tuning was completed once for each module in the camera in the lab and the resulting reference values were used for all data taking. 

It was discovered that temperature had a significant impact on these reference values resulting in an apparent increase in noise as the temperature of the modules decreased. This effect has been mitigated by implementing temperature-dependent trigger tuning (with a precision of 0.25$^{\circ}$ C). Each module FEE has four temperature sensors and the temperature of the modules is determined by averaging these four temperature readings. Each module is tuned at a variety of temperatures and the resulting reference values are stored in a database. 

During observations a five minute temperature equilibration time occurs prior to each data taking run. At the end of this time the temperature of each module is determined. From this temperature the correct reference values are found using the temperature dependent database and used during subsequent data taking.

\section{Camera Upgrade}

\begin{figure}
    \centering
    \begin{subfigure}{0.49\textwidth}
      \includegraphics[width=\textwidth]{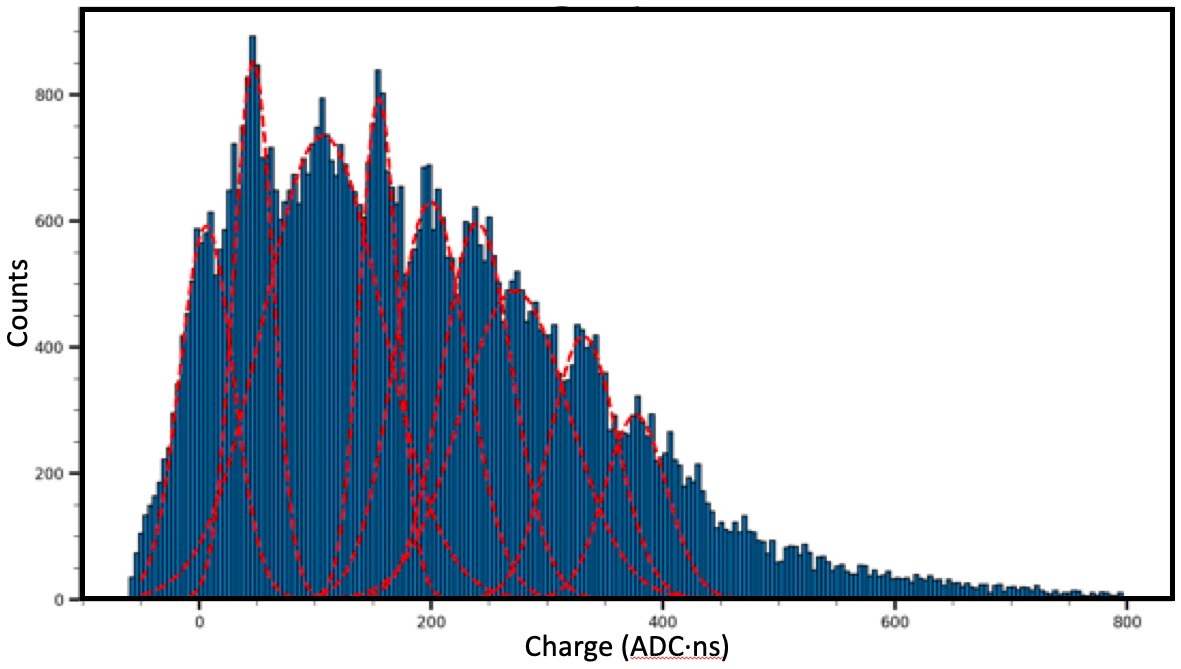}
      \caption{Charge spectrum of existing modules using Hamamatsu SiPMs and TARGET 7 electronics. An LED flasher was used to illuminate the entire module at once. A synchronous hardware trigger was used to trigger the module. \cite{Winter2017}}
    \end{subfigure}
    \hfill
    \begin{subfigure}{0.49\textwidth}
      \includegraphics[width=0.81\textwidth]{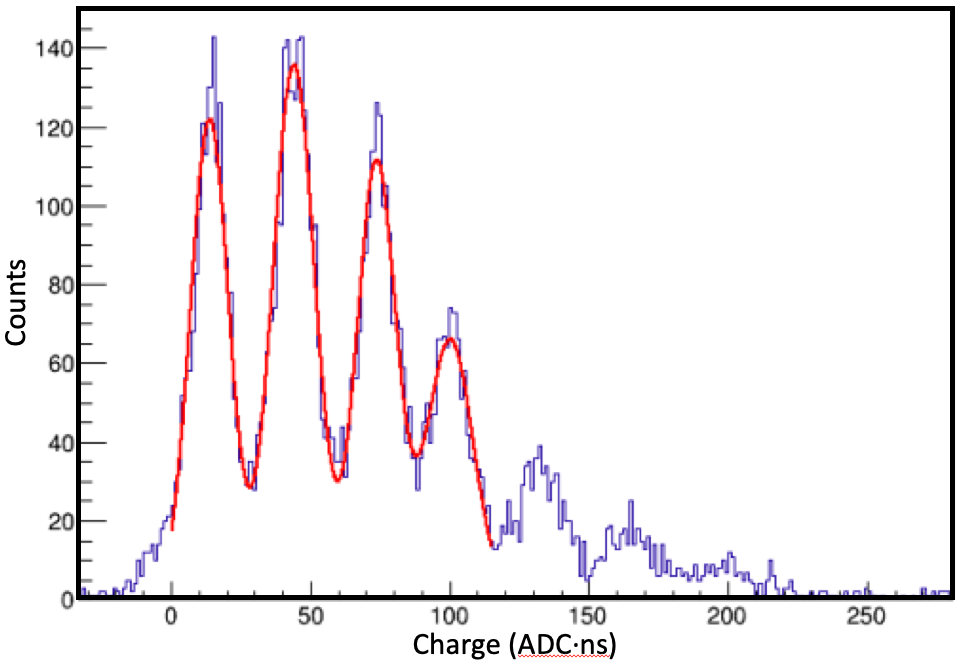}
      \caption{Charge spectrum produced from upgraded modules using FBK SiPMs and TARGETC electronics. Using a synchronous hardware trigger, a laser source combined with a moving stage was used to illuminate one pixel at a time.}
    \end{subfigure}
    \caption{The upgraded modules have lower electronic noise and therefore show a significant improvement in the charge spectrum resolution.}
    \label{fig:ChargeSpectrum}
\end{figure}

The camera upgrade will include an upgrade to both the SiPMs and front-end electronics. At present the camera uses a mixture of FBK NUV-HD and Hamamatsu SiPMs. After the upgrade the entire camera will use newer FBK NUV-HD SiPMs. These SiPMs will have improved photon detection efficiency, lower temperature dependence, and lower breakdown voltage than those modules currently used in the camera. The new FEE modules will incorporate next-generation TARGET C ASICs for digitization and sampling as well as a separate TARGET 5 Trigger Extension ASIC (T5TEA) for triggering. Figure \ref{fig:ChargeSpectrum} shows the charge distribution of the current and upgraded camera modules, illustrating a significant decrease in electronic noise. To connect the SiPMs to the FEE, a new custom ASIC called the SMART chip has been developed. The SMART chip will allow for improved SiPM bias voltage control as well as individual current measurements. It will also provide shaping and amplification of the analog signals. In addition to these upgrades the number of modules will also be increased from 25 to 177 in order to fully populate the focal plane. This will increase in the field of view from 2.7$^{\circ}$ square to 8.0$^{\circ}$ diameter.

In order to accommodate a fully populated focal plane the backplane PCB will also be redesigned. The current backplane was designed to accommodate 32 modules. The upgraded design will accommodate up to 25 modules and a total of nine backplanes will be installed in the camera (one for each sector). See Figure \ref{fig:CameraDesign}. DACQ boards will be mounted on top of their associated backplane. The heat management system will also be expanded in order to accommodate the increased number of modules in the camera. Additionally, the LED flashers will be upgraded, allowing for synchronous triggering with the backplanes and improved performance. \cite{meuresICRC2019}

The final SCT will include a Distributed Intelligent Array Trigger (DIAT) board which will provide GPS time-tagging as well as a synchronized 62.5~MHz clock to both backplanes and FEEs. \cite{2018NIMPA.891....6D} In addition to a DIAT board the pSCT upgrade will include the installation of a stand-alone time-tagging system. This system is divided into two modules: A laser-diode module located in the camera and a photo-detector/GPS module located near the camera server. A fiber optic communications cable will connect the two modules. Fast pulses (up to 5~kHz from camera backplane triggers) will be generated by a laser diode and then fed to a discriminator where a logic pulse is generated. A GPS time-stamp will then be attached to each pulse using custom-developed software/firmware. The system will provide stable timestamps for camera triggers to a GPS accuracy of 5~ns or better.

The pSCT camera upgrade has two main objectives: maximizing the camera's field of view, and upgrading the underlying electronics, enabling a lower trigger threshold and improved reconstruction and background rejection. The camera is scheduled to be upgraded by December of 2022.

\bibliographystyle{JHEP}
\bibliography{references}

\clearpage
\section*{Full Authors List: \Coll\ Consortium}

\scriptsize
\noindent
C.~B.~Adams$^{1}$,
G.~Ambrosi$^{2}$,
M.~Ambrosio$^{3}$,
C.~Aramo$^{3}$,
P.~I.~Batista$^{4}$,
W.~Benbow$^{5}$,
B.~Bertucci$^{2, 6}$,
E.~Bissaldi$^{7, 8}$,
M.~Bitossi$^{9}$,
A.~Boiano$^{3}$,
C.~Bonavolont\`a$^{3, 10}$,
R.~Bose$^{11}$,
A.~Brill$^{1}$,
A.~M.~Brown$^{12}$,
J.~H.~Buckley$^{11}$,
R.~A.~Cameron$^{13}$,
M.~Capasso$^{14}$,
M.~Caprai$^{2}$,
C.~E.~Covault$^{15}$,
D.~Depaoli$^{16, 17}$,
L.~Di~Venere$^{7, 8}$,
M.~Errando$^{11}$,
S.~Fegan$^{18}$,
Q.~Feng$^{14}$,
E.~Fiandrini$^{2, 19}$,
A.~Furniss$^{20}$,
A.~Gent$^{21}$,
N.~Giglietto$^{7, 8}$,
F.~Giordano$^{7, 8}$,
R.~Halliday$^{22}$,
O.~Hervet$^{23}$,
T.~B.~Humensky$^{24}$,
S.~Incardona$^{25, 26}$,
M.~Ionica$^{2}$,
W.~Jin$^{27}$,
D.~Kieda$^{28}$,
F.~Licciulli$^{8}$,
S.~Loporchio$^{7, 8}$,
G.~Marsella$^{25, 26}$,
V.~Masone$^{3}$,
K.~Meagher$^{21, 29}$,
T.~Meures$^{29}$,
B.~A.~W.~Mode$^{29}$,
S.~A.~I.~Mognet$^{30}$,
R.~Mukherjee$^{14}$,
A.~Okumura$^{31}$,
N.~Otte$^{21}$,
F.~R.~Pantaleo$^{7, 8}$,
R.~Paoletti$^{32, 9}$,
G.~Pareschi$^{33}$,
F.~Di~Pierro$^{17}$,
E.~Pueschel$^{4}$,
D.~Ribeiro$^{1}$,
L.~Riitano$^{29}$,
E.~Roache$^{5}$,
D.~Ross$^{34}$,
J.~Rousselle$^{35}$,
A.~Rugliancich$^{9}$,
M.~Santander$^{27}$,
R.~Shang$^{36}$,
L.~Stiaccini$^{32, 9}$,
H.~Tajima$^{31}$,
L.~P.~Taylor$^{29}$,
L.~Tosti$^{2}$,
G.~Tovmassian$^{37}$,
G.~Tripodo$^{25, 26}$,
V.~Vagelli$^{38, 2}$,
M.~Valentino$^{10, 3}$,
J.~Vandenbroucke$^{29}$,
V.~V.~Vassiliev$^{36}$,
J.~J.~Watson$^{39}$,
R.~White$^{40}$,
D.~A.~Williams$^{23}$,
A.~Zink$^{41}$,
\\
\\
\noindent
$^{1}$Physics Department, Columbia University, New York, NY 10027, USA.
$^{2}$INFN Sezione di Perugia, 06123 Perugia, Italy.
$^{3}$INFN Sezione di Napoli, 80126 Napoli, Italy.
$^{4}$Deutsches Elektronen-Synchrotron, Platanenallee 6, 15738 Zeuthen, Germany.
$^{5}$Center for Astrophysics | Harvard \& Smithsonian, Cambridge, MA 02138, USA.
$^{6}$Dipartimento di Fisica e Geologia dell'Universit\`a degli Studi di Perugia, 06123 Perugia, Italy.
$^{7}$Dipartimento Interateneo di Fisica dell'Universit\`a e del Politecnico di Bari, 70126 Bari, Italy.
$^{8}$INFN Sezione di Bari, 70125 Bari, Italy.
$^{9}$INFN Sezione di Pisa, 56127 Pisa, Italy.
$^{10}$CNR-ISASI, 80078 Pozzuoli, Italy.
$^{11}$Department of Physics, Washington University, St. Louis, MO 63130, USA.
$^{12}$Dept. of Physics and Centre for Advanced Instrumentation, Durham University, Durham DH1 3LE, United Kingdom.
$^{13}$Kavli Institute for Particle Astrophysics and Cosmology, SLAC National Accelerator Laboratory, Stanford University, Stanford, CA 94025, USA.
$^{14}$Department of Physics and Astronomy, Barnard College, Columbia University, NY 10027, USA.
$^{15}$Department of Physics, Case Western Reserve University, Cleveland, Ohio 44106, USA.
$^{16}$Dipartimento di Fisica dell'Universit\`a degli Studi di Torino, 10125 Torino, Italy.
$^{17}$INFN Sezione di Torino, 10125 Torino, Italy.
$^{18}$LLR/Ecole Polytechnique, Route de Saclay, 91128 Palaiseau Cedex, France.
$^{19}$Y.
$^{20}$Department of Physics, California State University - East Bay, Hayward, CA 94542, USA.
$^{21}$School of Physics \& Center for Relativistic Astrophysics, Georgia Institute of Technology, Atlanta, GA 30332-0430, USA.
$^{22}$Dept. of Physics and Astronomy, Michigan State University, East Lansing, MI 48824, USA.
$^{23}$Santa Cruz Institute for Particle Physics and Department of Physics, University of California, Santa Cruz, CA 95064, USA.
$^{24}$Science Department, SUNY Maritime College, Throggs Neck, NY 10465.
$^{25}$Dipartimento di Fisica e Chimica "E. Segr\`e", Universit\`a degli Studi di Palermo, via delle Scienze, 90128 Palermo, Italy.
$^{26}$INFN Sezione di Catania, 95123 Catania, Italy.
$^{27}$Department of Physics and Astronomy, University of Alabama, Tuscaloosa, AL 35487, USA.
$^{28}$Department of Physics and Astronomy, University of Utah, Salt Lake City, UT 84112, USA.
$^{29}$Department of Physics and Wisconsin IceCube Particle Astrophysics Center, University of Wisconsin, Madison, WI 53706, USA.
$^{30}$Pennsylvania State University, University Park, PA 16802, USA.
$^{31}$Institute for Space--Earth Environmental Research and Kobayashi--Maskawa Institute for the Origin of Particles and the Universe, Nagoya University, Nagoya 464-8601, Japan.
$^{32}$Dipartimento di Scienze Fisiche, della Terra e dell'Ambiente, Universit\`a degli Studi di Siena, 53100 Siena, Italy.
$^{33}$INAF - Osservatorio Astronomico di Brera, 20121 Milano/Merate, Italy.
$^{34}$Space Research Centre, University of Leicester, University Road, Leicester, LE1 7RH, United Kingdom.
$^{35}$Subaru Telescope NAOJ, Hilo HI 96720, USA.
$^{36}$Department of Physics and Astronomy, University of California, Los Angeles, CA 90095, USA.
$^{37}$Instituto de Astronom\'ia, Universidad Nacional Aut\'onoma de M\'exico, Ciudad de M\'exico, Mexico.
$^{38}$Agenzia Spaziale Italiana, 00133 Roma, Italy.
$^{39}$Deutsches Elektronen-Synchrotron, Platanenallee 6, D-15738 Zeuthen, Germany.
$^{40}$Max-Planck-Institut für Kernphysik, P.O. Box 103980, 69029 Heidelberg, Germany.
$^{41}$Friedrich-Alexander-Universit\"at Erlangen-N\"urnberg, Erlangen Centre for Astroparticle Physics, D 91058 Erlangen, Germany.

\end{document}